\documentstyle{article}
\begin{document}

\title{Pending problems in QSOs}

\author{Mart\'\i n L\'opez-Corredoira\\
Instituto de Astrof\'\i sica de Canarias,
E-38200 La Laguna, Tenerife, Spain\\
Departamento de Astrof\'\i sica, Universidad de La Laguna,
E-38206 La Laguna, Tenerife, Spain\\
E-mail: martinlc@iac.es}

\maketitle

{\bf \large ABSTRACT}

Quasars (Quasi Stellar Objects, abbreviated as QSOs) are still nowadays, 
close to half a century after their discovery, objects which are not completely understood. 
In this brief review a description of the pending problems, inconsistencies
and caveats in the QSO's research is presented. The standard paradigm model based on
the existence of very massive black holes that are responsible for the QSO's huge 
luminosities, resulting from to their cosmological redshifts, leaves 
many facts without explanation. There are several
observations which lack a clear explanation, for instance: 
the absence of bright QSOs at low redshifts, a mysterious evolution not 
properly understood; the inconsistencies of the absorption lines, such as 
the different structure of the clouds along
the QSO's line of sight and their tangential directions;
the spatial correlation between QSOs and galaxies; and many others.


\section{Introduction. QSOs: observations, standard view}

The topic of this article are the objects called 
Quasars (contraction of QUASi-stellAR radio source) 
or QSOs (Quasi-Stellar Objects). 
The first series of these objects 
were discovered with radio telescopes in the late 1950s. 
Many were recorded as radio sources with no corresponding visible object,
hence their name ``quasar'', but optical counterparts 
were discovered later.  Indeed, only around 10\% of these objects have 
strong radio emission ('radio-loud') (de Vries et al. 2006); hence 
it would be better to use the name 'QSO' to refer to them, 
including the 'radio-loud' and the 'radio-quiet' classes, although
the name ``quasar'' is also used for radio-quiet QSOs.

Their radiation is emitted across the spectrum
from X-rays to the far-infrared with a peak 
in the ultraviolet-optical bands, with some quasars also being 
strong sources of radio emission and of gamma-rays. 
Some quasars display rapid changes in luminosity in the 
optical and even more rapidly in the X-rays.
Their spectra is characterized by a high redshift and 
a combination of very broad lines, of several percent of the speed of light,
with narrow ``forbidden'' lines. 
Emission lines of hydrogen, mainly in the Lyman series and Balmer series, 
Helium, Carbon, Magnesium, Iron and Oxygen are the brightest lines. 
The atoms emitting these lines range from neutral ones to those having a degree of
ionization higher than the possible ionizations produced by star radiation.
In early optical images, quasars looked like point sources, 
indistinguishable from stars, except 
for their peculiar spectra. With infrared telescopes and the 
Hubble Space Telescope, the ``host galaxies'' surrounding the 
quasars have been identified in some cases.
  
According to the standard view, a QSO is an extremely powerful 
and distant active galactic nucleus.
Their redshift is believed to be cosmological, associated
to the expansion of the Universe, so their high redshift
implies long distances, and consequently, huge luminosities.
These objects are thought to be comprised of compact regions of
10-10,000 Schwarzschild radii across surrounding the central 
supermassive black hole of a galaxy.
The huge luminosity of quasars results from the accretion discs of 
central supermassive black holes, which can convert on the 
order of 10\% of the mass of an object into energy as compared 
to 0.7\% for the p-p chain nuclear fusion process that dominates 
the energy production in sun-like stars.
The widths of the broad lines resulting from Doppler shifts are due to 
the high speeds of the gas emitting those spectral lines. 
Fast motions strongly would indicate a large mass. 
Since they cannot continue to feed at high rates 
for 10 billion years, after the accretion of the surrounding gas 
and dust is terminated, they would become ordinary galaxies, in few tens of Myr.
Unified models were developed in which QSOs 
were classified as a particular kind of an active galaxy, like a Seyfert 1
but with higher luminosity due to the higher mass of the black hole 
and a general consensus emerged.  In many cases it 
is simply based in the viewing angle that distinguishes them from 
other classes of active galaxies, such as Seyfert 2, blazars, radio galaxies. 

There are plenty of books and reviews about these fascinating objects 
and the standard hypothesis to interpret them (e.g., Rees 1984;
Antonucci 1993; Kembhavi \& Narlikar 1999). In this review I pretend
to do something different: rather than presenting the successes of
the standard theory in our understanding of the QSOs, I want to show
the dark side, the aspects of which are still not very clear
and deserve further consideration, either for improving the present
standard theory, or to modify it, and even to change it completely if
it were necessary. This article is not a forum for the discussion
of all the possible theoretical approaches, but deals with the observational
facts which could affect the discussion of what is known and/or what is
still unknown.

Quasars (Quasi Stellar Objects, abbreviated as QSOs) are still nowadays, 
close to half a century after their discovery, objects which are not completely understood. 
In this brief review a description of the pending problems, inconsistencies
and caveats in the QSO's research is presented.  The standard paradigm model based on
the existence of very massive black holes that are responsible for the QSO's huge 
luminosities, resulting from their cosmological redshifts, leaves 
many facts without explanation. 
Possibly not all of the cited references of this review are correct. 
My task here is just to compile the bibliography on the pending 
problems, not to critically examine them.  This review is not complete, 
there are indeed hundreds or thousands of references
relevant to these questions, although I think the references hereby presented  
are quite representative.  Nonetheless,  given the sort of the material 
displayed in our references, one can get a general glimpse of what are the most relevant 
topics discussed nowadays pertaining to the nature of QSOs.

\section{Very high luminosity at high redshift}

As said previously,  the most remarkable characteristic of QSOs is perhaps their very high
luminosity. The luminosities of the brightest QSOs, using the standard 
interpretation, are as bright as several thousands of cD galaxies 
(the brightest galaxy in a cluster of galaxies at low redshift $z$). 
Only one QSO may be as bright as tens of 
large clusters of galaxies (with $\sim 1000$ large galaxies in each of them)
in a relatively compact region. The regions should be very compact
in order to justify their strong variability in short times.

To reduce their luminosity, some prefer to think that there
is a magnification due to gravitational lenses, but no evidence of that 
was found. Yamada et al. (2003) or Richards et al. (2004) examined 
the fields of some QSOs at $z\approx 6$ and concluded that they are not 
gravitationally magnified, so the luminosity at that redshift must be very high.
The anisotropy of the QSO radiation, observable only when the beam 
is pointed towards us, would also reduce the luminosity; but 
the number of sources would be much higher, a huge number, and we
would then see a large number of them in the nearby Universe, 
even if their beams of maximum flux are not pointing towards us; 
local AGNs (Active Galactic Nuclei) would experience also this
strong anisotropy radiation emission.  It is not the case, we do not
observe that.  Therefore, the luminosity of the QSOs must be really huge, 
provided that their assumed distance is correct.

Since their discovery, long debates have taken place on whether
the distance pointed by the redshift is real or not.
After that period of debate in the 60s and the early 70s, the mainstream of
astronomers adopted the consensus that the redshift of the
QSOs is cosmological in origin and therefore the luminosity is intrinsically very high. 
Hence, it could be explained in terms of supermassive black holes (e.g., Kembhavi \& Narlikar 1999, 
ch. 5; Djorgovski et al. 2008) while discarding other alternative
interpretations. Nonetheless, some apparent inconsistencies remain still 
within the standard explanation.

In the case of ultramassive black holes of around
$10^{10}$ M$_\odot $, necessary to explain the extremely high luminosities
of high redshift QSOs, 
they would attract the surrounding material at 
relativistic speeds and would become strongly redshifted (Kundt 2009), something which is not observed.
Some physical variables should be proportional to the distance of a source, such as the Faraday rotation or the time dilation factor, but they are not observed to be correlated to the redshift.
The polarization of radio emission rotates as it passes through magnetized 
extragalactic plasmas. Such Faraday rotations in QSOs should increase 
(on average) with distance. If redshift indicates distance, then rotation 
and redshift should increase together. However, the mean Faraday rotation 
is less near $z=2$ than near $z=1$ (Arp 1988).
Time dilation, which is observed in supernovae, should also be observed
in QSOs, increasing the periods of variability with the distance,
but it is not observed in QSOs against expectations (Hawkins 2010). 

Moreover, the huge dispersion in the magnitude-redshift relation for QSOs 
(Hewitt \& Burbidge 1987) makes impossible to derive a
Hubble law for them. This is not a strong argument since the intrinsic dispersion of luminosities might be high itself, but it might be 
pointing out that something is wrong with the distance measurement.

\section{Host galaxies}

The luminosity of the host galaxies,
which are supposed to be normal galaxies and whose luminosity come
only from the stellar emission, have also extremely high luminosities.
Schramm et al. (2008) detected host galaxies at $z=3$ 
which are extremely luminous: down to $M_{V,rest}=-26.4$, that
is, $L_{V,rest}\approx 3\times 10^{12}$ L$_{V,\odot }$.
Its color $(B-V)_{rest}\approx 0.0$ indicates a young population (0.3 Gyr), so
its stellar mass is $M_{*}\sim 5\times 10^{11}$ M$_\odot $ (Schramm et al. 2008),
somewhat high, although within the possible values.
Therefore, the explanation for these high luminosities is that we could be
observing a very young populations of stars. 

Magain et al. (2005) reported on the observation of a quasar lying at the edge 
of a gas cloud, whose size is comparable to that of a small galaxy, but 
whose spectrum shows no evidence for stars. The gas cloud is excited by the 
quasar itself. Magain et al. (2005) could not see any host galaxy in it;
if a host galaxy were present, it should be at least six times fainter 
than it would normally be expected to be for such a bright quasar. This tells us
that the host galaxies, although they are normally brighter than normal
galaxies, in some cases are much fainter or inexistent. 
We do not know the reason. 

Other problems to solve in the host galaxies remain: 
the dynamical mass of molecular gas
of a case at $z=6.4$ ($\sim 5.5\times 10^{10}$ M$_\odot $) is
too high to leave room for other kinds of matter, and it cannot
accommodate the predicted $10^{12}$ M$_\odot $ stellar bulge
necessary for its massive black hole (Walter et al. 2007).
There are also unexpected non-detections of cold neutral gas in 
the host galaxies of high redshift QSOs 
(at $>10^{23}$ W/Hz; Curran et al. 2008).  

\section{Age and metallicity of high redshift QSOs}

Some QSOs are apparently somewhat older than the Universe at their
corresponding redshift. For instance, the quasar APM 08279+5255 at redshift $z=3.91$ has an age of 2-3 Gyr, which
constrains $\Omega _m$ to be less than 0.21 (Jain \& Dev 2006),
lower than the accepted values for the standard cosmology nowadays.
Possibly the age measurement is somewhat overestimated and this
would explain the inconsistency, but it is important to bear in mind
that there are pending cases like this to be solved.

Big Bang requires that stars, QSOs and galaxies in the early universe 
be ``primitive'', meaning mostly metal-free, because it requires many generations 
of supernovae to build up metal content in stars. But the observations 
show the existence of even higher than solar metallicities 
in the ``earliest'' QSOs and galaxies (Fan et al. 2001, Becker et al. 2001,
Constantin et al. 2002, Simon et al. 2007). 
The iron to magnesium ratio increases at higher redshifts (Iwamuro et al. 2002).
And what is even more amazing: there is no evolution of some line ratios, 
including iron abundance (Dietrich et al. 2003, Freudling et al. 2003,
Maiolino et al. 2003, Barth et al. 2003) between $z=0$ and $z=6.5$, 
iron abundance at $z\sim 6$ QSOs is similar to its abundance in local QSOs.
The amount of dust in high redshift galaxies and QSOs is also much higher 
than expected (Dunne et al. 2003).
In view of these evidences, orthodox cosmologists claim now that the star 
formation began very early and produced metals up to the solar abundance quickly, in roughly half Gyr. 
However, it is not enough to come up with such a surprising claim, it needs 
to be demonstrated, and I do not see any evidence in favor of such a quick 
evolution in the local galaxies.

\section{Evolution or non-evolution of QSOs} 

There is another remarkable fact about the luminosity of QSOs. They are 
extremely bright at high redshift, 
but QSOs at low redshift have got a much lower
luminosity. From the analysis of the bolometric luminosity function of QSOs at
different redshift (Hopkins et al. 2007), it is clear that the relative abundance
of high luminosity QSOs decreases quickly at low redshift.
In visible, below $z=0.3$, the rate of luminosity decrease begins
to slow down and below $z=0.1$ the luminosity begins to increase again
(Bell 2007). At SDSS survey, all QSOs at $z<0.4$ are fainter than $M_B=-26$ 
(with K-corrections) while there are plenty of QSOs tens of times 
brighter than this limit at higher redshifts. 
Also in other wavelengths this fact is observed clearly:
in the X-ray region it is particularly strong the evolution at low redshift
(Shen et al. 2006); or in the radio regime (Bridle \& Perley 1984; Bell 2006, Figs. 9, 10).
A strong density and luminosity evolution is required.
It seems that we live in the era in which the bright QSOs have disappeared.

It is usual to claim that evolution is the wild card which solves this kind of
problems. Something very different should have happened at high redshift with
respect to the low redshift Universe to obtain this different level of luminosity.
However, no visible signs of this evolution are observed.
There is no indication of any significant evolution in the X-ray properties of 
quasars between redshifts 0 and 6, apart from the intrinsic luminosity, 
suggesting that the physical processes of accretion onto massive black 
holes have not changed over the bulk of cosmic 
time (Vignali et al. 2005). Also, the spectral features of low and high 
redshift QSOs are very similar (Segal \& Nicoll 1998).
There are not variations of black hole masses and Eddington ratios for equal luminosity QSOs (L\'opez-Corredoira \& Guti\'errez 2011).
Therefore, the situation is that QSOs have a strong evolution
in the values of their luminosity but not significant change in other properties, and it
is not well understood which is the cause of the luminosity evolution.
Possibly the environment might change. 
Nonetheless, do we know the connection between the triggering of activity and the 
environment?

\section{Triggering of activity}

It is usually suggested that the interactions with
the companion galaxies are related to the mechanism of feeding the  
black hole of the QSO (Stockton 1982; Canalizo \& Stockton 2001). 
Horst \& Duschl (2008) presented the results of an extremely simple 
cosmological model combined with an evolutionary scenario
in which both the formation of the black hole as well as the gas accretion 
onto it are triggered by major mergers of gas-rich galaxies. Despite the very 
generous number of approximations their model reproduces the quasar density evolution in remarkable agreement with some observations.
However, other authors (e.g., Kundt 2009) find difficulties 
to understand how the very massive black holes are formed. Kundt (2009) thinks
that centrifugal forces, pressures, and detonations prevent huge
amounts of material to be collapsed.
 
Many of the QSO host galaxies at low redshift
suffered mergers with accompanying starbursts (Jahnke et al. 2007;
Bennert et al. 2008a). Jahnke et al. (2007) 
showed that $\approx 50$\% of the host galaxies show distortions in their 
rotation curves or peculiar gas velocities above normal maximum velocities 
for disks, sign of mergers. And all host galaxies have quite young 
stellar populations, typically 1-2 Gyr. While this presents evidence 
for a connection of galaxy interaction and AGN activity for half of the sample, 
this is not clear for the other half. There is a $\approx 50$\% of the host
galaxies which are undistorted disk dominated.
Bennert et al. (2008b) think that most QSO host galaxies experienced 
mergers with accompanying starbursts but that the activity is triggered 
with a delay of several hundreds Myr after the merger.
Even so, why don't we see evidence of mergers in many host galaxies?

The relationship of AGN triggered by mergers is not so clear, 
there are many observations apparently pointing to the opposite direction. 
In fact, Coldwell \& Lambas (2006) have shown that 
quasars at $z<0.2$ systematically avoid high density regions,
living in regions less dense than cluster environments. The
environment of QSOs is populated by galaxies systematically bluer, 
and preferentially with disk-type morphology. And at $0.5\le z\le 0.8$, 
only 10\% of QSOs live in relatively rich clusters
of Abell richness class 1-2, and 45\% of them live in field-like
environments (Wold et al. 2001). AGNs live also in low-density regions
(Westoby et al. 2007; Constantin et al. 2008), in even lower density regions 
than QSOs (Strand et al. 2008); and QSOs with low mass black holes are
in lower density regions than those with high mass black holes
(Strand et al. 2008). Unless the velocity
in a cluster of galaxies is so high that strongly reduces the formation of mergers, the number of mergers should be higher in richer environments and it would lead to a lower triggering of activity and starbursts in low dense regions. We might also consider that galaxies in rich clusters are stripped of their interstellar medium by harrasment, so it would be reasonable that the QSO activity is less than in the field galaxies, but the ratio of spiral galaxies with non-stripped gas is still high enough to consider there should be activity triggering. More recently, Cisternas
et al. (2011) have shown directly that there is not an enhanced frequency of major merger signatures for the AGN hosts with respect to other galaxies, so this points out that major mergers should not be an important
element for the triggering of activity.

There is evidence for a significant post-starburst population 
in many luminous AGNs, and that a direct, causal link might exist between star 
formation and black hole accretion (Ho 2005).
The detection of large amounts of warm, extended, molecular gas 
also points that QSOs have vigorous star formation (Walter et al. 2007).
However, it is also common nowadays the proposal that 
AGN host galaxies are a transition population, being the AGNs 
the mechanism for star-formation quenching, where the black hole blows
out the gas.

Therefore, to sum up this section, we have no idea of the mechanism which
triggers the activity in galaxies, and the different observations point
to different directions within the actual proposed scenarios.

\section{Superluminal motions}

Superluminal motions of sources at high distance ($D$) 
are observed, i.e. angular speeds $\omega $ between 
two radio emitting blobs which imply linear velocities $v=D\omega $ greater
than the speed light (Cohen 1986). For instance,
the QSO 0805-077 presents apparent superluminal motions up to 
59.1 h$^{-1}$c (Lister et al. 2009).

There are some explanations.
The so called relativistic beaming model (Rees 1967) assumes that there is one blob
$A$ which is fixed while blob $B$ is traveling almost directly towards
the observer with speed $V<c$ with an angle $\cos ^{-1}(V/c)$ between
the line of motion and the line $B$-observer. This leads to an apparent
velocity of separation which may be greater than $c$. There is also another proposal 
in a gravitational bending scenario (Chitre \& Narlikar 1979). However,
both explanations share the common criticism of being contrived and
having somewhat low probability ($\sim 10^{-4}$) (Narlikar \& Chitre 1984).
In the case of blazars, the superluminal
motions in blazars can be statistically 
explained in the frame of the unification scheme of AGNs (Liu \& Zhang 2007). 

\section{Periodicity of redshifts}

Another problem with QSOs which has a long history and without
a clear agreement is the periodicity of redshifts.
In a homogeneous and isotropic universe we expect the redshift distribution
of extragalactic objects to approximate a continuous and
aperiodic distribution.  However, a periodicity with $\Delta z=0.031$ or 0.062
was found for the QSOs (Burbidge \& O'Dell 1972; Bell \& McDiarmid 2006;
Hartnett 2009), which cannot be understood in terms of the Cosmological Hypothesis.
Other authors found a periodicity of QSOs with a period of 0.089 the function
$log(1+z)$ instead of regular intervals linear in 
$z$ (Karlsson 1977; Napier 2006).

However, Hawkins et al. (2002), Tang \& Zhang (2005, 2010) 
found that there is no periodicity 
of QSOs in SDSS and 2dF surveys beyond randomness and selection effects.
Napier \& Burbidge (2003) argued that Hawkins et al. had
not measured the redshifts of these faint quasars with respect 
to the redshift of their active parent galaxies.
The periodicity found of the redshifts is measured with respect to the parent
galaxy; people who do not find a periodicity would simply measure the
redshift with respect to us ($z=0$)---say Napier \& Burbidge. Therefore, the
debate has not ended and it is not clear who is right.

\section{Correlation with galaxies of lower redshift}

There are several statistical analyses (e.g., Chu et al. 1984; Zhu \& Chu 1995; 
Burbidge et al. 1985; Burbidge 2001; Ben\'\i tez et al. 2001;
Gazta\~naga 2003; Nollenberg \& Williams 2005; 
Bukhmastova 2007) displaying an excess of high-redshift sources near the low-redshift 
galaxies, or positive and very significant cross correlations among surveys 
of galaxies and QSOs, or an excess of pairs of QSOs with very different redshifts, 
etc. 

There are plenty of individual cases of galaxies with an excess of QSOs with 
high redshifts near the centre of nearby galaxies, mostly AGNs 
(Arp 1987, 2003; Burbidge 2001; Bell 2002a,b; L\'opez-Corredoira \& 
Guti\'errez 2006). In some cases, 
QSOs are only a few arc-seconds away from the centre of the galaxies. 
Examples are NGC 613, NGC 1068, NGC 1097, NGC 3079, NGC 3842, NGC 6212, NGC 7541,
NGC 7319 (separation galaxy/QSO: 8"), 2237+0305 (separation galaxy/QSO 0.3"), 
3C 343.1 (separation galaxy/QSO: 0.25"), NEQ3, etc. 
In some cases there are even 
filaments/bridges/arms apparently connecting objects with different redshift: 
in NGC 4319+Mrk 205, Mrk273, QSO1327-206, NGC 3067+3C232 (in radio), NGC 622, 
NGC 3628 (in X-ray and radio), ESO 1327-2041+QSO 1327-206, 4C17.09, UGC 892,
NEQ3, etc. The probability of chance 
projections of background/foreground objects within a short distance of a 
galaxy or onto the filament is very low (down to 10$^{-8}$ or even lower). The 
alignment of sources with different redshifts also suggests that they may have 
a common origin, and that the direction of alignment is the direction of ejection. 
This happens with some configurations of QSOs around 1130+106, 3C212, 
NGC 4258, NGC 2639, NGC 4235, NGC 5985, GC 0248+430, etc. Another observation
suggesting the association QSO/galaxies with different redshift 
is that no absorption lines were found in QSOs corresponding to foreground 
galaxies (e.g. PKS 0454+036, PHL 1226), or distortions in the morphology of 
isolated galaxies.

The standard consensus is that all these cases are just random
projections of back\-ground/\-fore\-ground objects rather than the real associations
of objects with different redshifts. This might be true in many cases, but the
statistics still shows an excess number compared to the expected values for random projections. 
Hence, it remains difficult to explain these results in terms of random projections. 
Typical rebuffs such as ``it is an a posteriori statistical calculation'' or other considerations
such as a bias, incompleteness, gravitational lensing, do not solve the
anomalies in general (L\'opez-Corredoira \& Guti\'errez 2006; L\'opez-Corredoira 2010). 
On the other hand, the main supporters of the hypothesis 
of non-cosmological redshifts continue to produce tens of analyses of cases 
in favor of their ideas without too much care, pictures without rigorous 
statistical calculations in many cases, or with wrong identifications, 
underestimated probabilities, biases, use of incomplete surveys for 
statistics, etc., in many other cases. Some cases which were 
claimed to be anomalous in the past have found 
an explanation in standard terms (L\'opez-Corredoira 2010). 
There are, however, many papers 
in which no objections are found in the arguments and they present quite 
controversial objects, but due to the bad reputation of the topic, the 
community simply ignores them. This has become a topic in which everybody has an 
opinion without having read the papers or knowing the details of the problem, 
because some leading cosmologists have said it is bogus. 
Therefore, despite the many efforts by most cosmologists to forget this
old problem encountered with QSOs, the unexplained data are still there pointing out 
to us that we do not understand these phenomena completely. 
I maintain a neutral position, neither in favor of nor against 
non-cosmological redshifts. 

\section{Emission lines}

The standard model assumes that QSOs are the same type of objects as
Seyfert 1 galaxies but much brighter. Both of them present the characteristic
broad emission lines for hydrogen, carbon and other elements, 
together with a narrow emission of ``forbidden'' lines in the case of elements 
like oxygen, nitrogen, sulfur, etc. According to the standard black hole scenario and its accretion disc (e.g., Kembhavi \& Narlikar 1999,
ch. 5), the broad lines stem from the inner region with a strong velocity 
dispersion, while the forbidden lines would be generated in the outer regions.
These lines depend on the physical conditions of the clouds and the 
spectral energy distribution that photo-ionizes the clouds.
Although this scenario explains the main basic spectral features,
it remains to clarify some detailed observations. For instance,
some analyses of the spectra are given by Sulentic (2006), who believes that the double-peaked Balmer emission lines are better fitted with the bi-cone outflow model (Goldman \& Bahcall 1982, Robinson 1995) rather than the model of accretion disks, or perhaps a combination of accretion disks and outflows.
Indeed, the double-peaked optical emission lines are present only in $\sim 5$\%
of the AGNs, which raises another problem in supporting the black holes hypothesis as the engines of the activity.

According to the unification model, the differences between narrow-line AGNs
(Seyfert 2) and broad-line ones (Seyfert 1) stem from the orientation of
the toroidal regions of large extinction in the same type of galaxies. 
But the relative ratios of narrow and broad line AGNs cannot be 
reconciled with this simple model of unification in which the tori have the same optical depth 
and their opening angle is found to be independent of the luminosity (Lawrence 1987);
it requires a modification of the tori's extinction and/or cone angles 
for objects of different luminosity. 
Among the spiral AGNs with close satellite companions,
only a 2.6\% (1/39) of them are Seyfert 1 (Dultzin et al. 2008);
this cannot be explained either by a simple unification model and
requires modifications in terms of extra extinctions
in presence of companions.
Other observations,  pointing out that the differences between
Seyfert 2 and Seyfert 1 cannot be due entirely to different orientations of
the same object,  can also be found in the literature (e.g., Kembhavi \& Narlikar 1999,
\S 12.6.5).  Hence, it is required, in general,  a revision of the simple unification
model. This does not mean that the main aspects of the unification model
are wrong, but there are many observations which do not fit its predictions
unless the model is made more complex with the introduction of more ad hoc terms.

\section{Absorption lines}

According to the standard interpretation, the absorption lines in QSOs
are produced by the footprint in the intergalactic medium that the
paths of the photons take.  The redshift is associated to the 
path of the photons taken across the cloud. However, some observations 
are not always consistent with this scenario.
For instance, the HST NICMOS spectrograph has searched for objects
associated to the absorption lines of damped Ly-$\alpha $ systems (DLAs)
of some QSOs directly in the infrared, but failed for the 
most part to detect them (Colbert \& Malkan 2002). 
Moreover, the relative abundances of DLAs have a surprising uniformity, 
unexplained in the standard model (Prochaska \& Wolfe 2002),
except for the clouds which have a velocity difference less than 
6000 km/s from a QSOs, where the excess density by 
a factor 2 (at 3.5$\sigma$) (Russell et al. 2006) is something which
has neither a clear interpretation, 
unless the result by Russell et al.  is a random fluctuation (it is 
only 3.5$\sigma $).

The Ly-$\alpha $ forest is supposed to be produced by 
hydrogen in clouds of the intergalactic medium along the line of sight.
The temperature of these clouds does not change with
redshift (Zaldarriaga et al. 2001), a fact which does not fit the normal
predictions of the standard model since the density of the
clouds should have changed along the history of the Universe.
There is not clustering of the clouds or it is very weak (Dobrzycki et al. 2002), 
contrary to what it would be expected.

Hennawi \& Prochaska (2007) used a sample of 17 Lyman limit systems 
with column density N$_{HI}>10^{19}$ cm$^{-2}$ selected from 149 projected 
quasar pair sightlines, to investigate the clustering pattern of 
optically thick absorbers around luminous quasars at $z\approx 2.5$. 
Specifically, they measured the quasar-absorber correlation function 
in the transverse direction, and found a comoving correlation length 
of $r_0=9.2^{+1.5}_{-1.7}$ Mpc/h (comoving) assuming a power law correlation 
function with exponent $\gamma=1.6$. Applying this transverse clustering 
strength to the line-of-sight  would predict that $\sim $15-50\% of 
all quasars should show a N$_{HI}>10^{19}$ cm$^{-2}$ absorber within a velocity 
window of $v<3000$ Km/s. This value over-predicts the number of absorbers along 
the line-of-sight by a large factor, providing compelling evidence that 
the clustering pattern of optically thick absorbers around quasars is 
highly anisotropic. Hennawi \& Prochaska (2007) have argued that
the most plausible explanation for the anisotropy is that the transverse 
direction is less likely to be illuminated by ionizing photons than the 
line-of-sight direction, and that absorbers along the line-of-sight are being 
photo-evaporated.  An unbelievable explanation which serves to hide 
the fact that we have an unexpected observation within the 
standard interpretation of the origin of these absorption lines in QSOs.

Prochter et al. (2006) report on a survey for strong intervening MgII 
systems along the sight-lines of long-duration gamma-ray bursts (GRBs). 
The roughly four times higher incidence along the GRB sight-lines than in
QSOs sightlines with the same redshift is
inconsistent with a statistical fluctuation by greater than 99.9\% C.L. 
leading to a lower observed incidence along quasar sight-lines.
Prochter et al. (2006) analysis of these results  
suggest that at least one of our fundamental beliefs on 
the absorption-line research is flawed.
Tejos et al. (2009) confirmed it but with a factor three in the incidence
and only MgII systems having equivalent width at rest larger than 1 \AA .
However, Tejos et al. (2007) and Sudilovsky et al. (2007) 
conducted a similar study using CIV absorbers with GRB systems and their 
column density distribution and number density of this sample do 
not show any statistical differences with the same quantities measured 
in the QSO spectra. Maybe the discrepancy stems from a higher dust 
extinction in the strong MgII QSO samples studied up to now (Sudilovsky et 
al. 2007). Frank et al. (2007) propose that the solution is that the QSO beam
size is 2 times larger than the GRB beam sizes on average.
Porciani et al. (2007) have argued that the combined action of some effects 
can substantially reduce the statistical significance of the discrepancy.
Possibly this discrepancy can be solved in standard terms but it is not certain.

\section{Conclusions}

Irrespective of who is right or wrong,  either the researchers who are following the 
standard interpretation of QSOs or the others who are following a different one,  
the general impression which emerges from all of these problems is 
that we do not yet understand very well
many aspects of these objects and further research is still necessary. 

If I had to express my particular opinion, I would say
that the three most puzzling points are: 1) the total
absence of bright QSOs at low
redshift,  a mysterious evolution not properly understood;
2) the inconsistencies of the absorption lines, such as the different
structure of the clouds,  when performing comparisons between
measurements along the tangential and the line of sight of QSOs;
3) the spatial correlations among QSOs and nearby galaxies. 

Nonetheless, one must not forget that there are also good
reasons to support the standard scenario of QSOs, particularly
the results about the large distances and luminosities. Just to select
three among them: 1) the association of host galaxies with their QSOs
shows that the luminosity of the central part
of the object is much higher than the rest of the galaxy, and the hosts
have angular sizes decreasing with redshift; 2) the
absorption lines in many cases have a successful interpretation in terms
of gas or galaxies intervening along the line of sight; 3) cases involving gravitational
lensing indicate that the distance of
QSOs is much higher than the distance of the lensing galaxy. 
And, within the large
distance/luminosity assumption, there are also good reasons to support 
the standard paradigm model based on the existence of very massive black holes
with their accretion discs.

It would be desirable that we could proclaim
that we understand everything related to
these fascinating objects before other 50 years went by. 
But up to now we should leave at least 
some room for more discussions and even having
an open mind to embrace novel hypotheses in the interpretation of QSOs.

\

{\bf Acknowledgments:}
Thanks are given to Carlos Castro Perelman for proof-reading this paper.

\end{document}